\documentclass[sort=unsrt]{achemso}

\usepackage[version=3]{mhchem} 
\usepackage{
  amsmath,
  amssymb,
  array,
  booktabs,
  braket,
  color,
  float,
  graphicx,
  hyperref,
  ulem,
  comment
}

\title{Single-Contact Problem in Atomically Flat Interfaces: a Simulation Approach}

\author{Rui Dong}
\email{dongru@tcd.ie}
\author{Ahmed Uluca}
\author{Graham Cross}
\author{Stefano Sanvito} 
\affiliation{School of Physics, AMBER and CRANN Institute, Trinity College, Dublin 2, Ireland}

\begin{document}
\begin{abstract}
Understanding friction at single-asperity contacts is essential for bridging the gap between nanoscale 
structural superlubricity and realistic tribological systems dominated by Hertzian contact geometry. In 
this work, we combine atomistic simulations and a modified continuum model to investigate the onset 
of sliding at crystalline SiO$_2$/SiO$_2$ interfaces. Interfacial sliding potential energy surfaces (ISPES) 
are computed to determine the load-dependent shear strength and minimal-scale sliding (MSS) friction. 
Both quantities exhibit linear dependence on normal pressure below 3 GPa, and have non-zero values at zero pressure. Incorporating these 
parameters, we extend the classical Mindlin model by including adhesion and nanoscale load effects, 
allowing us to describe the stick to slip transition under realistic Hertzian stress distributions. The model shows that nonuniform pressure distributions substantially lower the effective static friction, and oscillatory-shear experiments on graphene-passivated contacts reproduce both the predicted stiffness-collapse signature and, in the passivated limit, the adhesion-limited shear strength obtained from simulation, supporting the model's relevance to real micro-asperity tribology.
\end{abstract}

\maketitle

\section{Introduction}

Nearly one-third of the energy produced by fossil fuels in vehicles is dissipated through frictional 
losses \cite{InfluenceofTribology}. Improving tribological technologies therefore represents a critical 
opportunity to reduce fuel consumption and mitigate CO$_2$ emissions. However, tribology has 
progressed relatively slowly compared to other materials-focused fields. This lag stems largely from 
the highly complex physical and chemical interactions at sliding interfaces, which are difficult to observe 
and model comprehensively by experiments under fully controlled conditions. To overcome this limitation, 
computational simulations offer a powerful means to address this challenge.

In recent years, structural superlubricity, the dramatic reduction of friction arising from incommensurate 
atomically flat interfaces, has emerged as a robust and well-studied phenomenon at the nanoscale 
\cite{HiranoShinjo, FirstExperimentMoS2,IncommCrystals,SSL_Review2025,Ying2025,Sun2024}. However, practical devices 
rarely present ideal, uniformly loaded, atomically flat contacts. Instead, contact mechanics are typically 
governed by micro asperities of Hertzian geometries that produce strongly nonuniform pressure and shear 
distributions \cite{Solhjoo2016,Ho2025}. Understanding how atomically flat interfaces behave under realistic Hertzian stress fields therefore represents 
a currently missing crucial bridge between proof-of-concept nanoscale superlubricity and scalable, device-level 
implementations.

Directly simulating even a microscopic Hertzian contact with atomistic precision is computationally expensive due to the large number of atoms involved and the complexity of local interfacial geometry and interactions \cite{Sharp2017,Wang2020,Luan2006}. Unfortunately, accurately capturing deformation, stress 
distribution, and energy dissipation requires high-resolution modeling, so that the need for significant computational resources can be hardly mitigated. Nevertheless, these simulations are essential to understanding the results arising from new experimental capabilities 
developed also to bridge the nano and macro scales.  

Significant efforts have been made to investigate the sliding behavior of nominally Hertzian contacts at the nanoscale. 
Luan {\it et al.} examined various adhesion models and demonstrated that the predicted results are highly sensitive 
to the specific geometry of the probing tip \cite{Luan2006}. Another important line of research has focused on 
dislocation-based models, which describe interfaces composed of two perfectly matched crystalline surfaces. In 
these models, interfacial sliding is governed by the nucleation and propagation of dislocation loops. The resulting 
interfacial shear strength is strongly dependent on the contact size and the elastic properties of the materials 
involved \cite{Hurtado1999, Sharp2017, Wang2020}. However, these simulations typically neglect adhesive interactions 
and generally assume a classical linear relationship between the local shear strength and the applied normal load.

Here, we address these challenges by proposing a modified continuum framework, parameterized by a small set of 
quantities derived from quasi-static atomistic simulations, which retains the essential physics of contact mechanics. 
This multiscale approach is used to describe the onset of sliding in crystalline SiO$_2$-SiO$_2$ contacts. It captures 
the load-dependent nanoscale friction while connecting the macro-scale sheared Hertzian-Mindlin behavior. 

In particular, our strategy proceeds as follows. We first calculate the interfacial sliding potential energy surface (ISPES) 
of flat-on-flat crystalline SiO$_2$ interfaces. From the ISPES, we extract the shear strength and a "minimal-scale 
sliding" (MSS) friction, and analyze their dependence on the normal load.\cite{Losi2023} These results are then incorporated into a 
modified Mindlin model to evaluate the total lateral force during small-amplitude oscillations before gross sliding. Our 
model successfully reproduced the effective lateral stiffness and its character under low and high normal loads. Most 
importantly, our results reveal that the Hertzian distribution of load and stress plays a beneficial role in the tribology of 
2D-like flat interfaces.

\section{Methodology}
Quasi-static simulations of the tribological interfaces are performed using the LAMMPS~\cite{LAMMPS} molecular 
dynamics package. To model covalent interactions within the SiO$_2$ slab, we train a machine-learning spectral 
neighbor analysis potential (SNAP)~\cite{SNAP}, using the dataset of Erhard et al.~\cite{Erhard2022}. This ensures 
an accurate representation of the silicon-oxygen interactions. Details of our potential are provided in the supplementary 
material (SM). The coupling between the SiO$_2$ slabs is described by a Lennard-Jones potential, with the $C_6$ 
coefficients extracted from van der Waals DFT calculations to ensure accuracy. \cite{Tkatchenko2009} The combination of SNAP and LJ potentials has been demonstrated to be effective for nanoscale tribology 
simulations.~\cite{Dong2023}

\section{Result}
\begin{figure*}[h!] %
\centering
\includegraphics[width=0.5\textwidth]{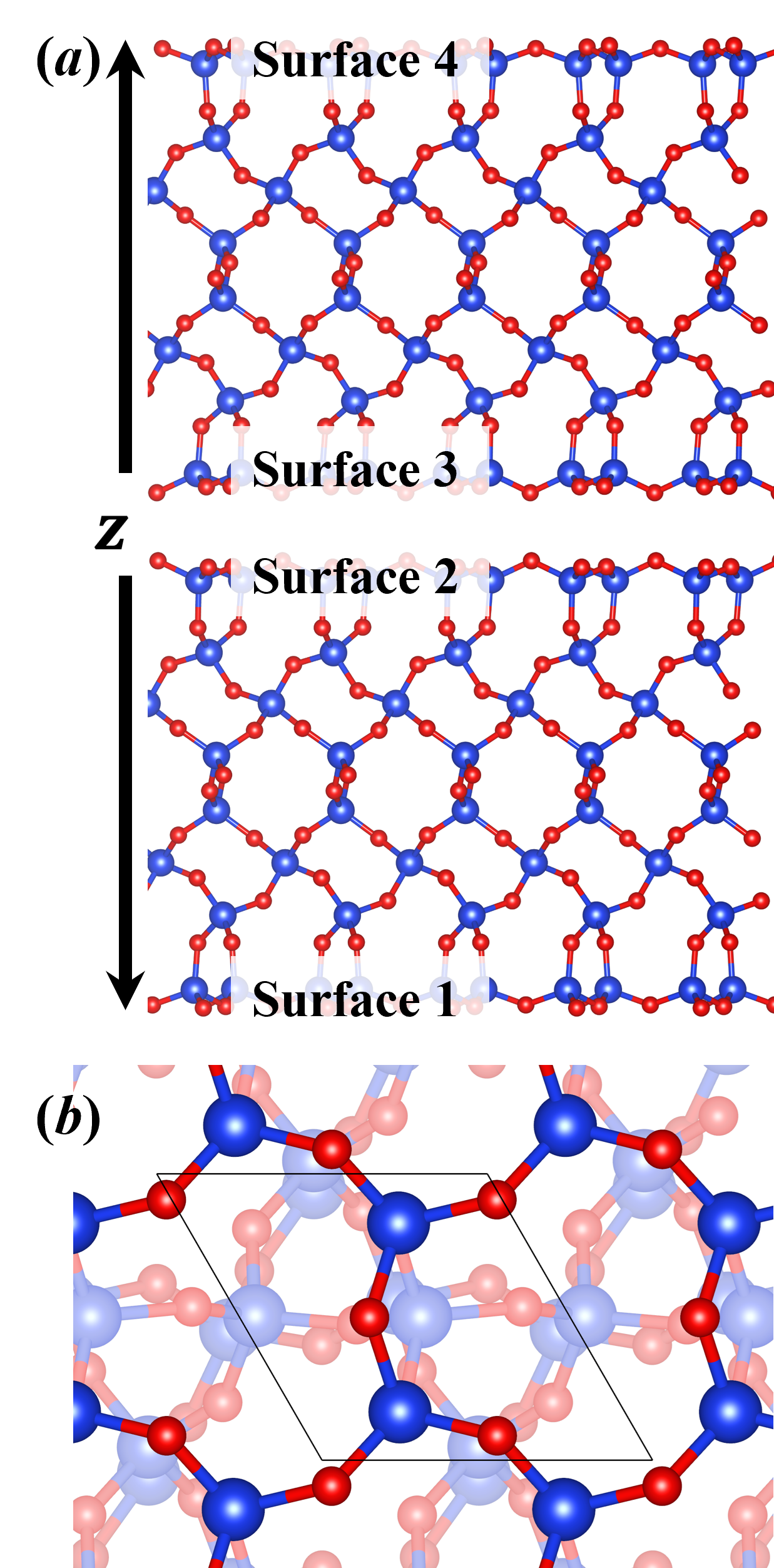}
\caption{The structure of the tribo-interface considered in this study. (a) Atomic structure of the SiO$_2$-SiO$_2$ 
flat-on-flat interfaces, and (b) top view of the ``dense'' surface of $\alpha$-quarts SiO$_2$. Highlighted in (b) is the 
six-ring reconstructed surface. Color code: silicon = blue, oxygen = red.}
\label{Fig1} 
\end{figure*}

We first simulate flat-on-flat SiO$_2$/SiO$_2$ interfaces to establish a local interfacial energy landscape baseline. 
Silica is well suited for tribological studies owing to its high hardness, chemical stability, and smooth 
surfaces~\cite{Li2011,Mate2011,Ciftci2026}. The interface consisted of two identical SiO$_2$ (001) slabs in the $\alpha$-quartz phase [see Figure~\ref{Fig1}(a)]. The body of each slab consists of 6 layers of O-Si-O and is sandwiched between two 
crystalline surfaces. The selected surface geometry is known as ``dense'' due to its high atomic 
density, see Fig.~\ref{Fig1}(b). This surface structure is chosen because it preserves stoichiometry, has no dangling bonds, and provides an atomically flat 
profile comparable to two-dimensional (2D) materials, therefore, greatly simplifies the simulation. In addition, the 
atomistically smooth surface prevents interfacial chemical bond-induced friction, as well as aging. This preserved 
2D-like geometry also rationalizes our choice in light of recent experimental studies 
\cite{imprintedRoughness2Ds,substrateEffect_Graphene,numberofLayersUniversalTrend} and our collaboration 
efforts with 2D materials, where single- and few-layer graphene serve as atomically thin passivation layers facilitating 
protected silicon-oxide sub-nm roughness keeping the interface two-dimensional of multi nano-asperities under a 
microscale Hertz contact between diamond and silicon oxide. 

\subsection{Interfacial sliding potential energy surface}
To construct the ISPES, the total thickness `$z$' [this is the distance between Surface 1 and 4 defined in Fig.~\ref{Fig1}(a)] 
is varied to impose different normal loads, while the relative lateral displacement $(x,y)$ of the upper slab is scanned across 
the primitive cell. The atoms in the surfaces 1 and 4 are held fixed after the initial optimization, while those on surface 2, 3 
and the inner atomic layers are fully relaxed, for every lateral displacement. A three dimensional matrix of total energy, 
$U_{ijk}=U(x_i,y_j,z_k)$, is thus obtained as a function of the spatial coordinates. Similarly, the normal load is also obtained as a matrix $p_{ijk}=p(x_i,y_j,z_k)$.

For a fixed value of $z=Z$, the surface $U(x,y,Z)$ corresponds to a constant volume ISPES, which does not directly 
map to a single normal load because of the atomic-scale corrugation. Alternatively, for a point $(X,Y)$, one can link $U(X,Y,z)$ and $p(X,Y,z)$ at the same $z$ to form a 1D relation:
\begin{equation}
\tilde{U}(p)|_{X,Y}.
\end{equation}
The energies at a series of target $P$ can be obtained using interpolation. The collection of energies at the same $P$ forms the constant pressure ISPES, $\tilde{U}(x,y)|_{P}$. Both the 
constant-volume and constant-pressure ISPES deviate from the expected perfect six-fold symmetry of the hexagonal 
lattices. This reflects the slight out-of-plane displacement of oxygen atoms, which breaks the ideal symmetry.

\begin{figure*}[h!] %
\centering
\includegraphics[width=1.0\textwidth]{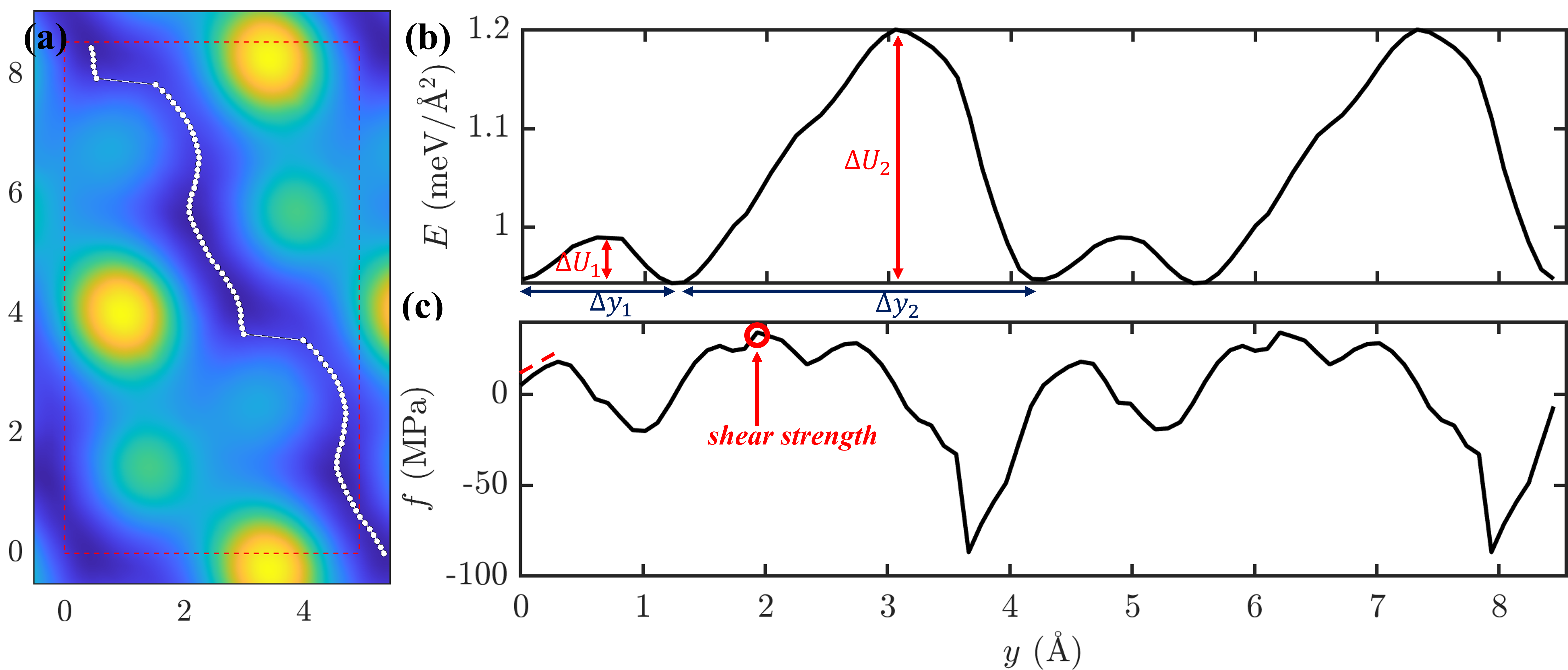}
\caption{Minimum energy pathway on the constant-pressure ISPES and the energy and force profiles at 0.5 GPa normal 
load. (a) The constant-pressure ISPES of SiO$_2$-SiO$_2$ flat-on-flat interface. The white line shows the minimum 
energy pathway and the red dotted lines denotes the rectangular 1$\times$2 supercell. Panel (b) shows the 
energy profile along the pathway. Zero energy is associated to the global energy minimum obtained at zero normal load. 
Red arrows mark the energy loss to overcome each barrier. The corresponding force along the $y$-axis of the pathway
is shown in (c). This is normalized to have the unit of stress. The dashed line shows the shear modulus for a tiny displacement 
and red circle marks the shear strength.}
\label{Fig2} 
\end{figure*}

The constant-pressure ISPES is subsequently analyzed to identify sliding pathways. The lateral force components, 
$f_x(x,y)$ and $f_y(x,y)$, are obtained from the numerical derivatives of $\tilde{U}(x,y)$. In our analysis, the pulling 
force is applied along the $y$-axis. As the top layer moves from one energy minimum to the next, the system consistently 
adjusts to positions of zero force along the $x$-axis, since no external force is applied in that direction. Figure~\ref{Fig2}(a) 
illustrates the ISPES at 0.5~GPa and the corresponding pathway with zero force along the $x$-axis. Owing to the broken 
symmetry mentioned earlier, the ISPES displays a canyon-like low-energy region that defines the preferred sliding path. 
Note that, although simulations are performed within the hexagonal primitive cell, post-processing steps, such as the construction 
of the constant pressure ISPES and the identification of sliding pathways, are carried out in the 1$\times$2 rectangular 
supercell for convenience.

The energy landscape along the minimum-energy pathway and the corresponding $y$-component of force are presented 
in Figure~\ref{Fig2}(b) and (c). Both quantities are normalized by area, as the interface is atomically flat and the relevant 
properties scale with the contact area. We choose meV/\AA$^2$ as the unit of energy $U$ and MPa as the unit of force 
(stress). The initial part of the force-displacement curve exhibits a linear response, with its slope corresponding to the 
interfacial shear modulus (\textit{G*})\cite{Rejhon2022,Bonfanti2017,Minkin2023}. The red circle in Figure~\ref{Fig2}(c) denotes the maximum force reached before the slab escapes from the initial energy 
minimum. Expressed in units of pressure, this maximum lateral force defines the shear strength, $\tau$, of the 
interface \cite{Minkin2023}.

Once the applied pulling force exceeds $\tau$, the slab begins to slide. Typically, it enters a motion mode named ``stick-slip'',
where the slab remains temporarily trapped in a local minimum before abruptly jumping to the next. Depending on the pulling 
force, velocity and the elastic coupling between the pulling stage and the slider, the slider may cross multiple barriers in a single 
jump. When the motion of slider is bounded such that it crosses only one energy barrier at a time, the quasi-static assumption 
is valid and the friction can be directly estimated from the energy dissipation. After overcoming one barrier, it enters the local 
minimum, and the energy lost contributes to the friction in full. The average friction per lattice period equals the total energy 
dissipated in overcoming the barriers divided by the travel distance. The definition is borrowed from the energy dissipation of stick-slip in the Prandtl-Tomlinson.\cite{Popov2012}. We name it as ``minimal-scale sliding'' (MSS), to distinguish this highly constrained, localized motion 
from gross sliding.
The frictional stress in MSS is calculated as
\begin{equation}
\sigma \equiv f_\mathrm{MSS}=\frac{\sum{\Delta{U}_i}}{\sum{\Delta{y}_i}},
\end{equation}
where $\Delta{U}_i$ is the height of each energy barrier and $\Delta{y}_i$ is the distance between local minima. 
The frictional stress also has the unit of pressure. As a convention we omit the subscript and use $\sigma$ instead 
of $f$ hereafter to refer to MSS friction. The previously mentioned dislocation-based models do not incorporate this 
frictional contribution. Such omission arises from their underlying framework, in which a dislocation loop, once nucleated 
at the contact edge, propagates toward the contact center. Under the quasi-static approximation, frictional resistance 
to dislocation motion is neglected. Consequently, although the terms stick and slip zones are retained, these models 
do not exhibit true partial slip behavior; instead, the terminology is used solely to denote regions separated by the locus 
of maximum local stress. In the present work, we explicitly retain frictional effects, as our objective is to more closely mimic 
experimental conditions, where contact surfaces do not have perfect geometries and atomic or lattice misalignment is 
inevitable.

\begin{figure*}[h!] %
\centering
\includegraphics[width=1.0\textwidth]{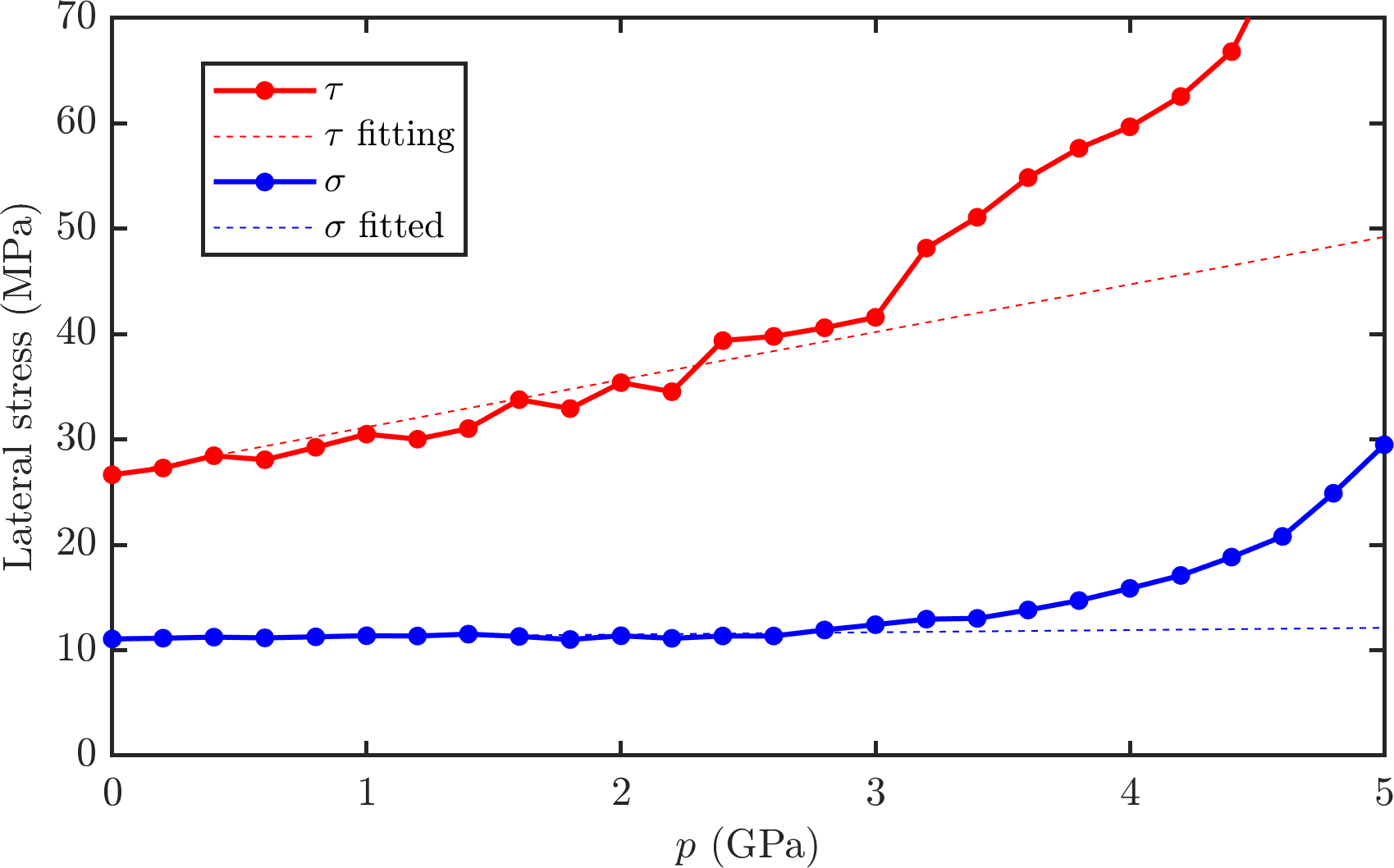}
\caption{The shear strength, $\tau$, and the coefficient of friction in minimal-scale sliding, $\sigma$, are plotted 
against the normal load, $p$, together with their linear fit. Here both $\tau$ and $\sigma$ are extracted from 
the constant-pressure ISPES. Both quantities have a linear dependence on the external load in the low pressure 
limit.}
\label{Fig3} 
\end{figure*}

A series of constant pressure ISPES are constructed under various normal loads, from which the shear strength 
$\tau$ and MSS friction $\sigma$ are extracted, see Figure~\ref{Fig3}. Both $\tau$ and $\sigma$ exhibit a nearly 
linear dependence on the normal load in the low-pressure regime ($p < 3$ GPa). Beyond this threshold, they increase 
sharply, coinciding with the onset load of subsurface plastic deformation in silica. Notably, MSS friction is less than half of the shear strength, indicating that the lateral force decreases once gross sliding begins. {This trend is consistent with the well-established kinetic-static friction gap.\cite{Popov2012}

The dependence of the shear strength and MSS friction on the load is well described by linear fits of the form 
\begin{equation}
\tau=k{\cdot}p+\tau_0\:,
\end{equation}
and
\begin{equation}
\sigma=\mu{\cdot}p+\sigma_0\:.
\label{EQF} 
\end{equation}
Notably, both $\tau$ and $\sigma$ retain finite values at zero external load ($\tau_0$ and $\sigma_0$, respectively) in 
contrast to the classical macroscopic view that friction vanishes without normal pressure. This behavior arises from 
the adhesion between atomically flat surfaces, which sustains friction even in the absence of external load. Moreover, 
the increase of $\tau$ and $\sigma$ with load is modest compared to their finite intercepts, underscoring the dominant 
role of interfacial adhesion at the nanoscale. The fitted parameters are $k$ = 4.52 MPa/GPa, $\tau_0$ = 26.7 MPa, 
$\mu$ = 0.21 MPa/GPa, and $\sigma_0$ = 11.1 MPa.

\subsection{Modified Mindlin model}
Under a uniform distribution of normal and lateral stress across the contact area, the extracted values of $\tau$ and 
$\sigma$ would directly correspond to the maximum static friction and the gross sliding friction, respectively. In this 
scenario, partial slip does not occur and the entire interface simultaneously transitions from stick to slip. However, such 
uniform loading is rarely realized in practical contacts formed by micro asperities. Instead, Hertzian pressure distributions 
are far more common. It is therefore important to examine the behavior of crystalline SiO$_2$/SiO$_2$ interface under 
Hertzian stress distributions.

The classical Mindlin model is widely used to analyze incipient sliding under Hertzian contacts \cite{Johnson1983}. 
In the original formulation, static and dynamic friction are assumed to share the same coefficient, a constant independent 
from the load, and both vanish in the absence of applied normal load. While effective at the macroscopic scale, this treatment 
neglects nanoscale phenomena such as adhesion and load-dependent shear strength. In the present work, we introduce 
a modified Mindlin model that incorporates the parameters extracted from our ISPES calculations. This extension provides 
a more realistic description of 2D-like interfaces under non-uniform Hertzian stress distributions.

For Hertzian contact between a spherical tip and a flat substrate, the contact radius, $a$, is given by
\begin{equation}
a = \frac{{\pi}Rp_0}{2E^*},
\end{equation}
where $E^*$ is the effective Young's modulus, $R$ is the radius of the tip, and $p_0$ the maximum normal pressure 
at the center of the contact\cite{Johnson1983}. In our analysis, we keep $R$ = 5 $\mu$m to be consistent with our experimental 
setup, and $E^*$ = 52.5 GPa, obtained from our DFT calculation of bulk SiO$_2$ elastic constants.

The radial distribution of Hertzian normal pressure is expressed as
\begin{equation}
p(r) = p_0\sqrt{1-r^2/a^2},
\end{equation}
with the maximum pressure, \textbf{$p_0$}, at the center and vanishing at the contact edge. The average normal pressure, 
$\bar{p}$, over the contacting area is $2p_0/3$, therefore the total applied load ($P$) on a Hertz contact of area $\pi*a^2$ is
\begin{equation}
    P=\dfrac{2}{3}\pi a^2p_0
\end{equation}
By combining this pressure distribution with the load-dependent shear strength obtained from our ISPES simulations, we 
calculate the local shear strength, $\tau(r)$ varying with local normal pressure, $p(r)$ shown in Figure~\ref{Fig4}(a) for a 
value of $p_0 = 1$ GPa. As a notable departure from conventional Mindlin analysis, the presence of van der Waals adhesion 
forces in our simulation means that $\tau(r)$ does not vanish at the contact edge. This is quantified by the finite intercept 
$\tau_0$ in Eq.~(\ref{EQF}), which would remain zero in pure Hertzian contact. Our modified Mindlin model maintains the 
principle of local slip where an applied lateral stress exceeds the local shear strength. 

In the classical Mindlin (Hertz) model, if a full-stick condition of no relative interfacial motion (i.e. no partial slip) is assumed, 
then all the lateral strains occur equally/uniformly, while the normal stress is non-uniformly distributed with respect to the 
Hertz equations. If a uniform lateral displacement $u$ is applied to the contact, the distribution of shear stress is given by
\begin{equation}
q(r) = \frac{q_0}{\sqrt{1-r^2/a^2}}\:,
\end{equation}
where the peak shear stress at the center $q_0$ is defined as   
\begin{equation}
q_0 = \frac{G^*u}{{\pi}a}\:,
\end{equation}
and $G^*$ = 29 GPa is the effective shear modulus obtained from our DFT calculations. This distribution predicts a minimum 
shear stress at the center and diverges at the edge (singularity). It means that a full stick condition is not physical because 
partial slip always initiates at the contact edge, where the imposed shear stress exceeds the local shear strength. We 
numerically determine the critical radius $c$, where $q(c) = \tau(c)$. This defines the boundary between the central stick 
zone and the outer slip zone. It is important to note that the coexistence of stick and slip zones in incipient sliding under 
Hertzian contact differs conceptually from this stick-slip motion mentioned in previous section. Figure~\ref{Fig4}(a) also 
shows the full stick shear stress of $u$ = 0.2 nm as an example. The intersection of $q(r)$ with $\tau(r)$ identifies the 
critical radius $c$.

Within the stick zone ($r < c$), the distribution of lateral shear stress remains unchanged. The contribution to the total 
lateral force can be calculated as an integral of the shear stress from 0 to $c$, namely
\begin{equation}
Q_\mathrm{stick} = 2\pi\int_{0}^{c}{\frac{q_0}{\sqrt{1-r^2/a^2}}rdr}\:.
\end{equation}
In the slip zone ($c < r < a$), slip occurs. However, since the tip as a whole has not yet entered gross sliding, relative 
motion remains minimal and follows the stick-slip type. The MSS friction is then applicable; the contribution to the total 
lateral force is therefore calculated by integrating the MSS friction from $c$ to $a$ (by the Coulomb relation),
\begin{equation}
Q_\mathrm{slip} = 2\pi\int_{c}^{a}{({\mu}p_0\sqrt{1-r^2/a^2}+\sigma_0)rdr}\:,
\end{equation}
where the coefficients $\mu$ and $\sigma_0$ are from Eq.~(\ref{EQF}). The total lateral force is then given by the sum 
of the two contributions, $Q = Q_\mathrm{stick} + Q_\mathrm{slip}$.

\begin{figure*}[h!] %
\centering
\includegraphics[width=0.5\textwidth]{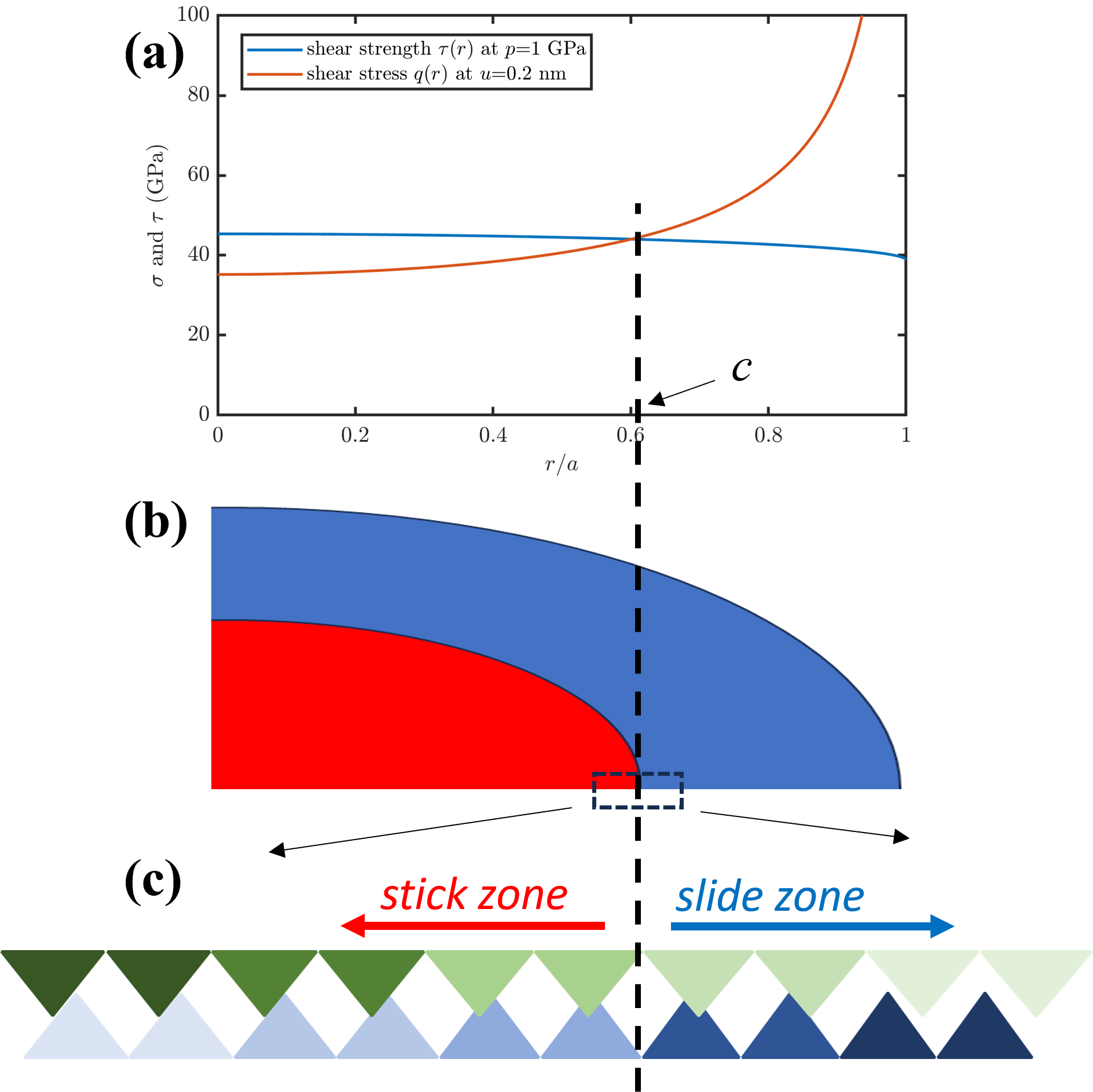}
\caption{Demonstration of the modified Mindlin model for atomically flat contacts. (a) For SiO$_2$-SiO$_2$ contact 
at 1GPa, the radial distribution of shear strength and shear stress of fully stick contact with lateral displacement of 0.2 nm. 
The crossover shows the boundary of stick zone. (b) Demonstration of the stick zone (red) and sliding zone (blue). 
(c) Demonstration of atomic scale asperities (atomic level corrugation) at the boundary.}
\label{Fig4} 
\end{figure*}

In principle, adhesive contact models such as DMT\cite{DMT}, JKR\cite{JKR}, or the Maugis--Dugdale transition model\cite{MDModel} should be considered. These models are primarily used to correct the contact radius and account for pull-off forces. Aside from minor deviations near the contact edge, the distribution of normal stress within the contact radius still follows the classical Hertzian contact theory. Therefore, adopting an adhesive contact model does not alter the results or conclusions of the above discussion. Given the relatively high elastic modulus of SiO$_2$, we adopt the DMT model to estimate the error in contact radius arising from adhesion. Using the parameters in this study, the contact radius predicted by the DMT model is approximately 30\% larger at 0.5 GPa, but the difference rapidly decreases to below 5\% when the pressure exceeds 1.2 GPa. 

Another source of error at small loads is the deviation from the continuum model. This effect becomes negligible at higher pressures. For example, using the parameters in this study, a pressure of 1 GPa corresponds to a contact radius of $\sim$150 nm, equivalent to approximately $\sim$330 lattice spacings. This value is well above the typical breakdown limit of the continuum model, which is on the order of 100-200 lattice spacings.\cite{Luan2005}

The modified Mindlin model establishes a direct relationship between total lateral force, $Q(u)$, and lateral 
displacement $u$ under given normal loads $\bar{p}$, see Figure~\ref{Fig5}. As the displacement increases, 
two competing effects arise within the stick zone: the shear stress continues to increase, but the size of the 
stick region decreases. Consequently, the stick contribution increases initially, reaches a maximum, and then 
decreases. This mechanism gives rise to the characteristic indenter displacement overshoot observed in displacement 
controlled experiments. In contrast, the weaker slip contribution increases monotonically with lateral displacement. 
Once the entire contact area is converted into a slip zone, the lateral force saturates, and the system enters gross sliding.

If the shear strength is relatively small compared to the MSS friction, the overshoot feature disappears when the saturated 
frictional force exceeds the peak of the stick contribution. The original Mindlin model, which has the same coefficient of 
shear strength and friction, is well below the limit for overshoot. It is also important to distinguish between driving modes: 
under displacement control, the full $Q(u)$ curve can be traced, including the decline beyond $Q_{\max}$, analogous to 
a spring stretched beyond its elastic limit. Under load control, however, gross sliding is triggered earlier at $Q_{\max}$ 
rather than at the saturation force.

\begin{figure*}[h!] %
\centering
\includegraphics[width=1.0\textwidth]{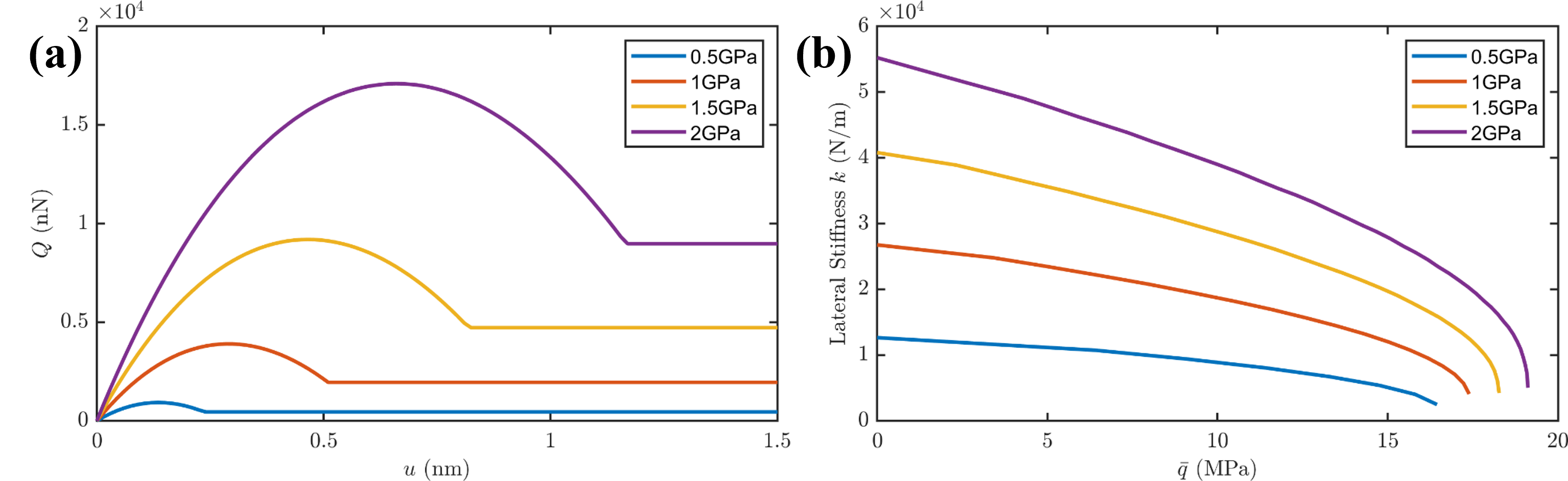}
\caption{SiO$_2$-SiO$_2$ interface sliding characteristics obtained from the modified Mindlin model at various normal 
pressures. (a) Total lateral force versus sliding distance of single strike sliding. When the lateral displacement increases, 
the lateral stress in the stick zone increases causing the stick zone to shrink. Total lateral force reaches a maximum then 
dropping to a constant sliding force. (b) Effective lateral stiffness in oscillation at various normal loads. One assumes the 
tip does lateral oscillation under load control and the force-displacement relation follows data in (a). Average lateral stress 
is calculated for oscillations with different mean displacement. See text for details.}
\label{Fig5} 
\end{figure*}

The $Q(u)$ relationship described above applies to a one-direction, single incipient-sliding event. In the oscillatory-shear experiments reported in Section~\ref{sec:exp} (Figure~\ref{fig:Fig6}), however, the tip is driven by a sinusoidal lateral force (load control), $Q(t) = Q_{\max}\sin(2\pi \nu t)$, and the corresponding 
displacement amplitude is measured. The driving frequency ($\nu$=80 Hz) is several orders of magnitude lower than the typical
phonon frequencies, validating the quasi-static approximation. We therefore assume a one-to-one mapping between force 
and displacement throughout a quarter oscillation cycle until the maximum force is reached, such that $u(t) = u(Q(t))$. From 
this assumption, the average displacement can be computed for each oscillation amplitude:
\begin{equation}
{\langle}u{\rangle}^2 = \frac{2}{\pi}\int_{0}^{\pi/2}{u^2[Q_{max}sin(\theta)]d{\theta}}\:.
\end{equation}
Then, we obtain the relation between the average force and average displacement, ${\langle}Q{\rangle}\sim{\langle}u{\rangle}$. 
The relation corresponds to a series of oscillations with different $Q_{\max}$. The effective lateral stiffness of the contact is then 
obtained as the derivative $k = d\langle Q \rangle / d\langle u \rangle$. The resulting stiffness-stress curves, $k(\langle q \rangle)$, 
are presented in Figure~\ref{Fig5}(b).

When $Q_{\max}$ reaches the maximum of the $Q(u)$ curve, any further displacement reduces the lateral response 
of the interface. Under load control, the applied force does not decrease, and this mismatch triggers the gross sliding. 
From this point, the quasi-static assumption breaks. The effective lateral stiffness consequently drops to zero, and the 
corresponding mean lateral shear stress $\langle q \rangle = \langle Q \rangle / \pi a^2$ can be interpreted as the 
effective shear strength, namely the maximum static friction. With increasing normal load, the effective lateral stiffness 
curve develops a pronounced knee-like feature just before vanishing due to the sudden transition of oscillatory shearing 
to the reciprocating sliding conditions, a behavior consistent with experimental observations-reference to the next section.

\section{Experimental support from passivated single-asperity contacts}
\label{sec:exp}
To assess whether the idealized picture developed above captures the essential
phenomenology of a real single-asperity contact, we performed oscillatory-shear
experiments with a diamond spherecone ($R = 5~\mu$m) on silicon oxide. These
measurements are intended as a conceptual, order-of-magnitude consistency check
on the model rather than a direct test of the simulated crystalline
SiO$_2$/SiO$_2$ interface. The simulated system is a clean, atomically flat,
adhesion-limited 2D-like contact, a limit that a bare oxide surface does not
realize: silica--silica and diamond--silica junctions are chemically metastable,
and their measured friction is dominated by silanol (Si--OH) molecular bridges
forming across the interface rather than by the conservative interfacial
potential captured in our ISPES.\cite{Zilibotti_Passivation, CUTINI2023601_GenericDiaSilica, Peng2023}
To approach the idealized limit experimentally, we introduce pristine
mechanically exfoliated few-layer graphene on the silicon oxide, which chemically
passivates the sub-nm nano-asperities and protects them from wear, thereby
presenting an ensemble of as-near-as-possible 2D-atomic contacts of the kind
treated in the simulation.

Figure~\ref{fig:Fig6} contrasts the two cases at a representative peak Hertz
pressure of $p_0 = 1.77$~GPa. For both contacts the lateral stiffness is tracked
as the oscillatory shear-force amplitude is ramped up, and the mean shear stress
is obtained by dividing the applied lateral force amplitude by the ideal Hertz
contact area. In each case the stiffness collapses sharply once the mean shear
stress reaches the effective shear strength, marking the
oscillatory-to-gross-sliding transition predicted by the modified Mindlin model.
The bare 300~nm SiO$_x$ contact collapses near $\sim$145~MPa, whereas the
graphene-passivated contact collapses near $\sim$26.7~MPa. Overlaid are modified
Mindlin curves evaluated at $\tau_0 = 145$~MPa and $\tau_0 = 26.7$~MPa,
respectively; the latter is the zero-load intercept obtained directly from our
ISPES simulations, applied here with no further adjustment.

\begin{figure}[h!]
    \centering
    \includegraphics[width=0.85\linewidth]{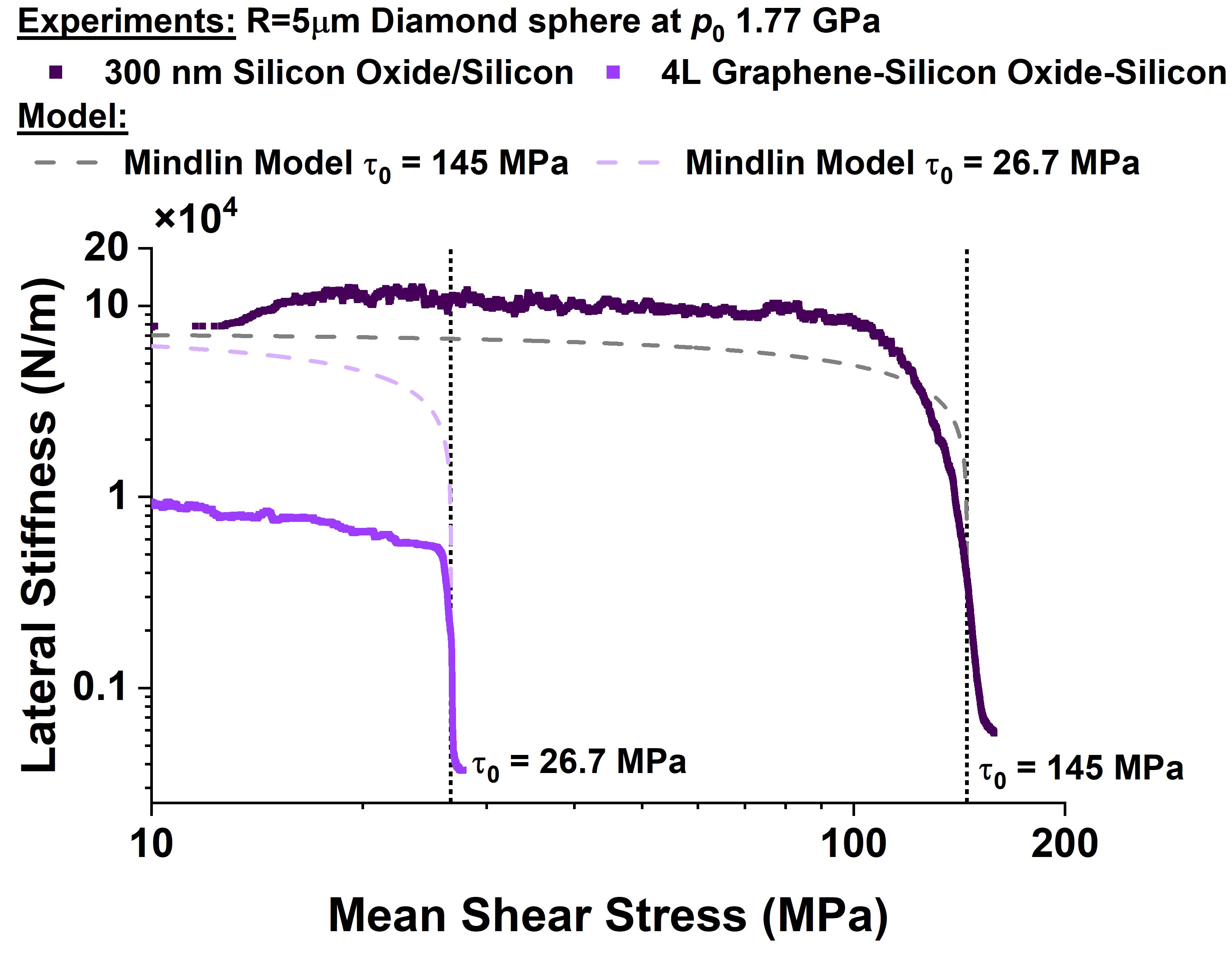}
    \caption{Oscillatory-shear response of a diamond spherecone ($R = 5~\mu$m) on
    bare versus graphene-passivated silicon oxide at $p_0 = 1.77$~GPa. Lateral
    stiffness, measured at 80~Hz by a lock-in amplifier, is plotted against mean
    shear stress (applied lateral oscillation force amplitude divided by the ideal
    Hertz contact area). Dark symbols: bare 300~nm SiO$_x$/Si; light symbols:
    4-layer graphene/SiO$_x$/Si. Dashed lines are the modified Mindlin model
    evaluated at $\tau_0 = 145$~MPa and $\tau_0 = 26.7$~MPa. The passivated
    contact collapses close to the simulated intercept $\tau_0 = 26.7$~MPa, while
    the bare contact collapses at a roughly five-fold higher stress consistent
    with silanol-bridge-dominated adhesion.}
    \label{fig:Fig6}
\end{figure}
Two features support the model. First, both contacts reproduce the characteristic
knee-like collapse of lateral stiffness at a well-defined critical mean shear
stress, the signature of the oscillatory-to-reciprocating transition that the
modified Mindlin model generates once the mean shear stress exceeds the effective
shear strength. Second, and more tellingly, chemically passivating the interface
toward the simulated 2D-like limit shifts the transition down to $\sim$26.7~MPa,
in close agreement with the adhesion-limited intercept $\tau_0$ obtained
independently from the ISPES. The bare contact, by contrast, requires a five-fold
larger $\tau_0$ that lies outside the scope of the conservative model and is
attributable to interfacial silanol chemistry. We therefore read this agreement
as evidence that the model captures the correct phenomenology and the correct
order of magnitude in the passivated, adhesion-limited regime it is built to
describe, not as a quantitative validation of the bare SiO$_2$/SiO$_2$ junction.
The full normal-load dependence of the transition, measured across several
locations and applied loads, is consistent with this picture and is provided in
the Supplementary Material. The measured transition is additionally affected by
roughness, non-ideal sphericity, contact relaxation, and junction growth; a
detailed treatment of these effects is deferred to forthcoming publications.

\section{Discussion}
The modified Mindlin model enables the direct calculation of tribological properties under nanoscale Hertzian contact 
using only two inputs: the load-dependent shear strength and the MSS friction obtained from ISPES. The analysis 
reveals that the Hertzian pressure distribution lowers the overall effective shear strength (static friction) compared to 
the uniform pressure distribution. This reduction arises since the distribution of normal pressure produces lower local 
shear strengths near the contact edge, where distribution of shear stress shows a higher value. As a result, local relative 
motion starts from the edges. Within the slip zone, the lateral force switches to MSS friction, which is approximately 
one-quarter of the shear strength. The emergence of partial slip therefore enables a gradual onset of sliding, substantially 
reducing the effective static friction compared with the uniform-load case. Quantitatively, the predicted effective shear strength in the passivated, adhesion-limited regime is on the order of tens of MPa, consistent with the $\sim$26.7~MPa transition measured on graphene-passivated contacts (Figure~\ref{fig:Fig6}). Reported values for bare single-asperity oxide and ceramic contacts are considerably higher, exceeding 200~MPa~\cite{BRAZIL2021106776}, in line with the $\sim$145~MPa transition we observe on bare SiO$_x$. We attribute this difference not to a failure of the contact-mechanical model but to interfacial chemistry: bare junctions form silanol molecular bridges that raise the effective adhesion well above the conservative interfacial potential captured by the ISPES. The passivation experiment thus isolates the regime in which the present model applies, and the agreement of the simulated intercept $\tau_0$ with the passivated transition supports this interpretation.

\section{Conclusion}

In this study, we have combined atomistic simulations with a modified continuum model to investigate nanoscale 
friction at crystalline SiO$_2$/SiO$_2$ interfaces. By calculating the interfacial sliding potential energy surface, we 
have extracted both the shear strength and minimal-scale sliding (MSS) friction as functions of applied normal load. 
The results reveal linear load dependence in the low-pressure regime, followed by a sharp increase beyond 3~GPa. 
Importantly, the MSS friction is consistently smaller than the shear strength, reflecting the well-known difference 
between static and dynamic friction observed at larger scales. 

Building upon these microscopic insights, we developed a modified Mindlin model that incorporates finite adhesion 
and the load dependence, thereby extending the classical framework to the nanoscale. This model reproduces the essential experimental phenomenology of passivated SiO$_2$ contacts, including the abrupt collapse of lateral stiffness at the onset of gross sliding, and recovers the adhesion-limited shear strength in the passivated limit. Our analysis demonstrates that the 
Hertzian distribution of stress reduces the effective static friction by promoting edge-initiated partial slip.

\bibliography{ref}

\end{document}